\documentclass[10pt,a4paper]{article}
\usepackage[utf8]{inputenc}
\usepackage[T1]{fontenc}
\usepackage{amsmath}
\usepackage{amssymb}
\usepackage{makeidx}
\usepackage{graphicx}
\usepackage{caption}
\usepackage{subcaption}
\usepackage{amsfonts}

\begin{document}

\title{Primordial dust universe in the Hořava–Lifshitz theory}

\author{G. Oliveira-Neto and A. Oliveira Castro Júnior\\
Departamento de F\'{\i}sica, \\
Instituto de Ci\^{e}ncias Exatas, \\ 
Universidade Federal de Juiz de Fora,\\
CEP 36036-330 - Juiz de Fora, MG, Brazil.\\
gilneto@fisica.ufjf.br, alessandroocj@protonmail.com
\and G. A. Monerat\\
Departamento de Modelagem Computacional, \\
Instituto Polit\'{e}cnico, \\
Universidade do Estado do Rio de Janeiro, \\
CEP 28.625-570, Nova Friburgo - RJ - Brazil.\\
monerat@uerj.br}

\maketitle

\begin{abstract}
We apply quantum cosmology to investigate the early moments of a Friedmann–Lemaître–Robertson–Walker (FLRW) cosmological model, 
using Ho\v{r}ava-Lifshitz (HL) as the gravitational theory. The matter content of the model is a dust perfect fluid. 
We start studying the classical model. Then, we write the total Hamiltonian of the model, quantize it and find the appropriate Wheeler-DeWitt equation. In order to avoid factor ordering ambiguities, in the Wheeler-DeWitt equation, we introduce a canonical transformation. We solve that equation using the Wentzel–Kramers–Brillouin (WKB) approximation and compute the tunneling probabilities for the birth of that universe ($TP_{WKB}$). Since the WKB wavefunction depends on the dust energy and the free coupling constants coming from the HL theory, we compute the behavior of $TP_{WKB}$ as a function of all these quantities. 
\end{abstract}



\section{Introduction}

The first work studying a cosmological model by means of quantum cosmology (QC) appeared in 1967 and was due to Bryce DeWitt \cite{dewitt}. The main motivation was studying the initial moments of the Universe and trying to remove the initial {\it big bang} singularity, present in the classical cosmological models. In that first work, 
DeWitt derived the fundamental equation of QC, known as Wheeler-DeWitt equation \cite{dewitt,wheeler}. One of the most interesting mechanisms which may explain the origin of our Universe without the presence of an initial {\it big bang} is the tunneling mechanism. It is one of the several properties of a quantum mechanical system which greatly differs from any property of a classical system. That mechanism implies that a quantum system can pass through a barrier that would be impenetrable by the corresponding classical system. Using the tunneling mechanism QC explains the birth of the universe by saying that it tunneled through a potential barrier from the origin of space and time to a later configuration with a finite size \cite{grishchuk,vilenkin,vilenkin1,vilenkin2,hawking,linde,rubakov,vilenkin3}. Many works have been published using that important idea \cite{rubakov1,vilenkin4,paulo1,acacio,germano,salvador,gasperini,germano1,germano2,rocha,vilenkin5,gil4}.

Petr Hořava introduced a new geometrical theory of gravity in 2009 \cite{horava} in an effort to overcome some obstacles found in General Relativity (GR). The main interest behind this was to obtain a renormalizable and unitary theory of gravity. This theory, nowadays called \textit{Hořava-Lifshitz theory} (HL), is known for its use of an anisotropic scaling between space and time found in condensed-matter systems. This anisotropy is measured through a dynamical critical exponent $z$. In this framework, Lorentz symmetry is not considered as fundamental. Instead, it is assumed as absent at high energies but it appears as an emergent symmetry at long distances. That is, at low energies, HL theory tends to GR and Lorentz symmetry is restored. Hořava built his theory using elements from ADM formalism \cite{misner}. 
In this formalism, the four-dimensional spacetime can be understood as being composed of stacked layers. These layers represent spacelike surfaces measured at different values of the time variable. Due to that, the spacetime is described through quantities coming from the decomposition of the four-dimensional spacetime metric $g_{\alpha\beta}$. The metric $g_{\alpha\beta}$ is separated into a three-dimensional metric $h_{ij}$, for the spacelike sections, the \textit{shift vector} $N_i$ and the \textit{lapse function} $N$. The most general situation is the one where all these quantities should depend on both space and time. In an attempt to make things easier, Hořava made two assumptions: the first one was to impose that $N$ should depend only on the time variable, which is known as \textit{projectable condition}. The second assumption was to reduce the number of terms contributing to the potential component of his theory, which is known as \textit{detailed balance condition}. The detailed balance condition was not a good choice though. It was shown that this version of the Hořava-Lifshitz theory has massive ghosts and instabilities \cite{mattvisser1}. Since its introduction, many works have emerged investigating different aspects of Hořava-Lifshitz theory. Some of them have shown interesting results, as one can verify in \cite{bertolami,kord,pedram2,wang1,gil3,gil,gil2,cordero,vicente,gao3,compean2,gil5}.

In this letter, we apply quantum cosmology to investigate the early moments of a FLRW cosmological model, 
using HL as the gravitational theory. The matter content of the model is a dust perfect fluid and the spatial sections have positive curvature.
Regarding the HL theory, we consider the {\it projectable} version of that theory without the {\it detailed balance condition}.
We start studying the classical model and obtain its total Hamiltonian. 
From that total Hamiltonian, we draw the phase portrait of the model, which gives some information on the scale factor behavior. Then, we 
quantize the model and find the appropriate Wheeler-DeWitt equation. In order to avoid factor ordering ambiguities, in the Wheeler-DeWitt equation, we introduce a canonical transformation. We solve that equation using the WKB approximation and compute the tunneling probabilities for the birth of that universe ($TP_{WKB}$). Since the WKB wavefunction depends on the dust energy and the free coupling constants coming from the HL theory, we compute the behavior of $TP_{WKB}$ as a function of all these quantities.

\section{The classical model and a canonical transformation}

In the present letter, we want to study the initial moments of a cosmological FLRW model in HL gravitational theory. The matter content
of the model is a dust perfect fluid and the spatial sections have positive curvatures. In order to carry out that
study, we use quantum cosmology. Therefore, we must start writing the total Hamiltonian of the theory, which is divided in two sectors: the gravitational sector, due to the HL theory, and the matter sector, due to the dust perfect fluid. Here, we follow the authors of Ref. \cite{kord}, using the ADM formalism \cite{misner} in order to determine the gravitational sector of total Hamiltonian and the Schutz variational formalism \cite{schutz,schutz1} in order to determine the matter sector of the total Hamiltonian. After all the calculations the result is \cite{kord},
\begin{equation}
    H_{T} = N\mathcal{H}_{T} = N \left[ -\frac{p_a^2}{4a} - g_c a + g_\Lambda a^3 + \frac{g_r}{a} + \frac{g_s}{a^3} + p_T\right].
\label{hamiltonian}
\end{equation}
The phase space of that model is described by the scale factor $a$, a variable associated to the fluid $T$ \cite{rubakov1} and their canonically conjugated momenta $p_a$ and $p_T$, respectively. It is important to mention that $p_T > 0$, because it is related to the dust energy density. All of them are functions only of the time coordinate $t$. The coefficients
$g_c$, $g_\Lambda$, $g_r$ and $g_s$ are coupling constants introduced by the HL theory \cite{maeda}, such that $g_c > 0$.
$H_T$ Eq. (\ref{hamiltonian}) is restricted to the situation where the FLRW spatial sections have positive curvatures
and the perfect fluid is dust.

Observing $H_T$ Eq. (\ref{hamiltonian}), we notice that due to the term $-p_a^2/4a$, when one quantizes $H_T$ it will develop factor ordering ambiguities. Then, it is important to try removing any possible factor ordering ambiguities, already, at the classical level. One may do that by performing a canonical transformation. We start that process doing the following redefinitions:
$N = \frac{\Bar{N}}{3}$, $\Bar{g_c} = \frac{g_c}{3}$, $\Bar{g_\Lambda} = \frac{g_\Lambda}{3}$, $\Bar{g_r} = \frac{g_r}{3}$, $\Bar{g_s} = \frac{g_s}{3}$ and $\Bar{p_T} = \frac{p_T}{3}$. Then, the total Hamiltonian (\ref{hamiltonian}) becomes,
\begin{equation}
    H_{T} = N\mathcal{H}_{T} = \Bar{N} \left[ -\frac{p_a^2}{12a} - \Bar{g_c} a + \Bar{g_\Lambda} a^3 + \frac{\Bar{g_r}}{a} + \frac{\Bar{g_s}}{a^3} + \frac{\Bar{p_T}}{a^{3\omega}}\right].
\label{hamiltonian-2}
\end{equation}
\noindent Next, we perform the following canonical transformation \cite{nelson},
\begin{equation}
    a = \left( \frac{3x}{2} \right)^{2/3},\ \ \ p_a = p_x a^{1/2},
\label{variable_change}
\end{equation}
\noindent With the aid of Eqs. (\ref{variable_change}), we may write $H_T$ Eq. (\ref{hamiltonian-2}) as, 
\begin{equation}
    H_{T} = \frac{p_x^2}{12} + V(x) - \Bar{p_T},
\label{hamiltonian_new}
\end{equation}
where
\begin{equation}
    V(x) = \Bar{g_c} \left( \frac{3x}{2} \right)^{2/3} - \Bar{g_\Lambda} \left( \frac{3x}{2} \right)^{2} - \Bar{g_r}\left( \frac{3x}{2} \right)^{-2/3} - \Bar{g_s}\left( \frac{3x}{2} \right)^{-2},
\label{potential}
\end{equation}
and we are working in the gauge where $\Bar{N}=-1$.

\begin{figure}[!h]
  \centering
	\includegraphics[width=0.45\textwidth]{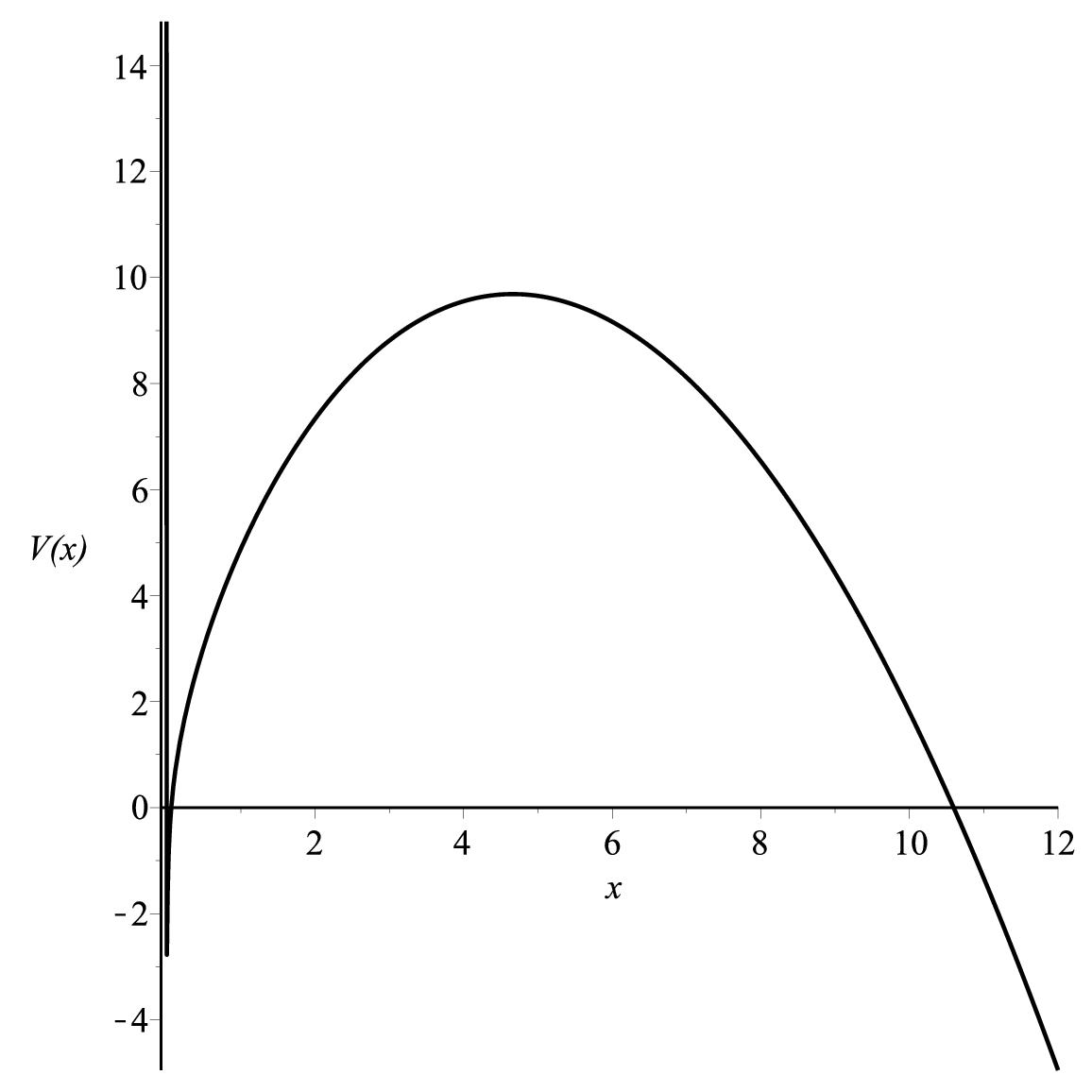}
	\caption{Potential barrier obtained from $V(x)$ where $\Bar{g_c} = 4.0, \Bar{g_\Lambda} = 0.1, \Bar{g_r} = 0.18$ and $\Bar{g_s} = -0.0001$.}
\label{potential-1}
\end{figure}

Looking for well defined potential barriers, we will work with models which have $\Bar{g_r} \geq 0$, $\Bar{g_\Lambda} > 0$ and $\Bar{g_s} < 0$. Under these conditions, the most general potential barrier obtained from $V(x)$ Eq. (\ref{potential}) has the shape of an infinite tall barrier at the origin, followed by a well, followed by a second barrier. The choice of $\Bar{g_s} < 0$ implies that the potential has the infinite tall barrier at the origin and those models are free from the big bang singularity, at the classical level. We show in Figure \ref{potential-1} an example of the most general potential barrier obtained from $V(x)$. 

\begin{figure}[!tbp]
  \centering
  \includegraphics[width=0.45\textwidth]{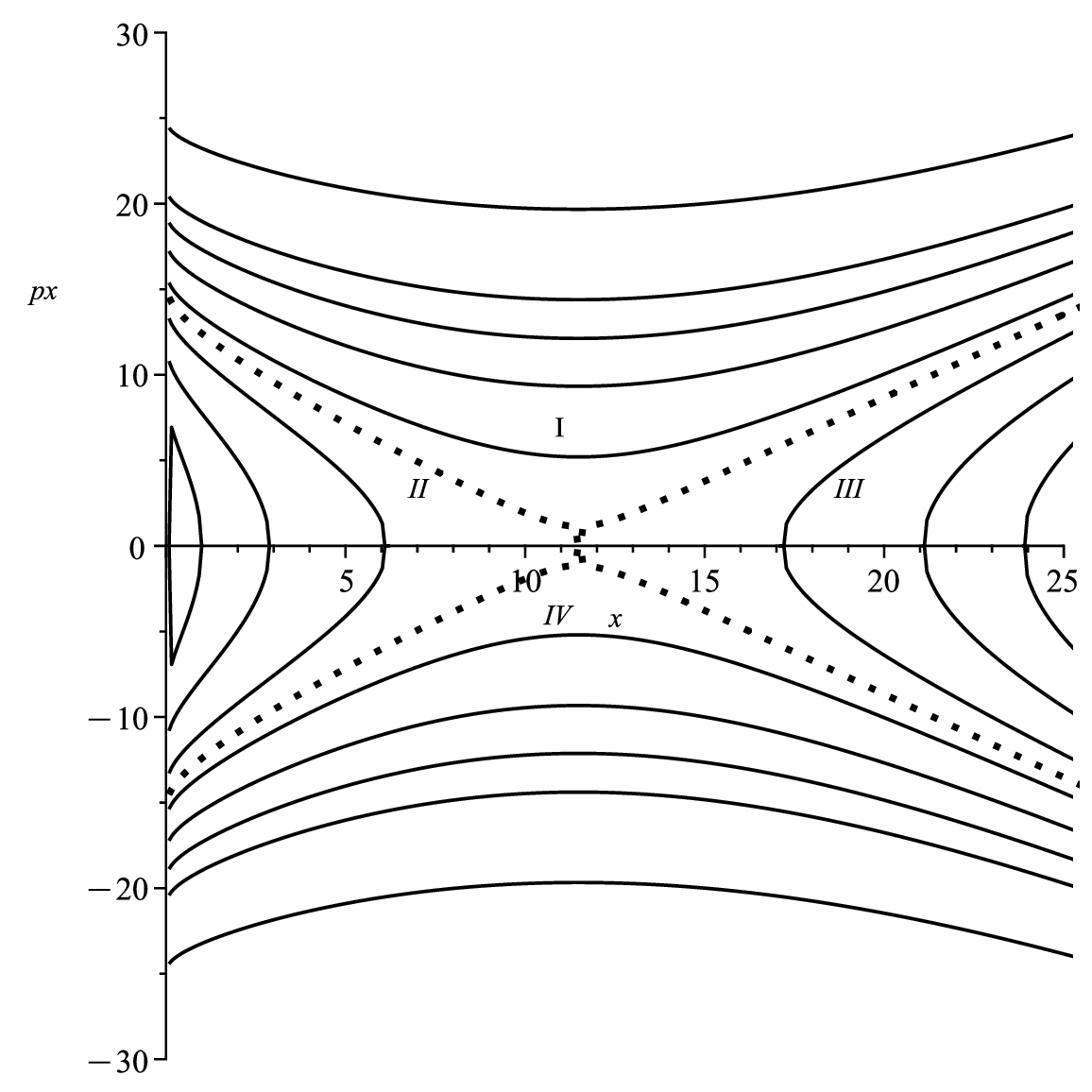}
  \caption{Phase portrait for the model where the potential has the shape of an infinite tall barrier close to the origin, followed by a well, followed by a second barrier. The parameter $\Bar{p_T}$ varies from $0$ to $300$. The dotted lines are called separatrices because they separate different types of solutions.}
\label{phase_portrait}
\end{figure}

In order to understand a little bit more about this model, we draw its phase portrait. We do that by imposing the constraint equation $H_T=0$, where $H_T$ is given in Eq. (\ref{hamiltonian_new}). We show an example in Figure \ref{phase_portrait}. In this phase portrait, the values of the HL's coupling constants are given by: $\Bar{g_c} = 4.0, \Bar{g_\Lambda} = 0.03, \Bar{g_r} = 0.2$ and $\Bar{g_s} = -0.001$. From the phase portrait Figure \ref{phase_portrait}, it is possible to identify qualitatively four different types of classical solutions. 
These solutions are: (i) A periodic solution which has a scale factor that starts expanding from an initial small value and then continue expanding. When it reaches a maximum value, it begins a contraction. Once it reaches a minimum value, it starts expanding again to repeat this cycle. This kind of solution is found at region II. 
(ii) A bouncing solution where the scale factor begins at a large value and contracts until it reaches a minimum value, where it starts an expansion at an accelerated rate to infinity. This kind of solution is found at region III. 
(iii) An expansive solution where the scale factor begins at a small value and grows, at an accelerated rate, towards infinity. This kind of solution is found at region I. 
(iv) A solution where the scale factor starts contracting from a large value and continue contracting to a small value. This type of solution appears in region IV.

\section{Quantum model and WKB tunneling probabilities}

Next, we want to write the quantum cosmology version of the present model, described by $H_T$ Eq. (\ref{hamiltonian_new}). In order to do that, we apply the Dirac's formalism for quantization of constrained systems \cite{paulo2}, which produces the following Wheeler-DeWitt equation,
\begin{equation}
    \left( -\frac{1}{12}\frac{\partial^2}{\partial x^2} + V(x) \right)\Psi(x,\tau) = i \frac{\partial \Psi(x,\tau)}{\partial \tau},
\label{WdW}
\end{equation}
where $V(x)$ being the one from equation (\ref{potential}), $\Psi(x,\tau)$ is the wavefunction of the Universe and the new variable $\tau$ is defined by, $\tau = -T$. As anticipated that equation is free from factor ordering ambiguities. Assuming that (\ref{WdW}) has a solution like $\Psi(x,\tau) = \psi(x)e^{-iE\tau}$, one gets,
\begin{equation}
    \left[ \frac{d^2}{dx^2} + 12(E-V(x)) \right]\psi(x) = 0,
\label{WdW-2}
\end{equation}
where $E$ is the energy associated to the dust perfect fluid.

In order to solve Eq. (\ref{WdW-2}), we use the WKB approximation \cite{merzbacher}. With the aid of that solution, we compute the transmission coefficient, or tunneling probability, which will be written as $TP_{WKB}$. After all the calculations, we find \cite{merzbacher},
\begin{equation}
 TP_{WKB} = \frac{4}{\left( 2\theta + \frac{1}{2\theta}\right)^2}.
\label{tpwkb}
\end{equation}
For the present model $\theta$ is given by,
\begin{equation}
    \theta = \exp\left( \int_{x_l}^{x_r} \sqrt{12\left( \Bar{g_c} \left( \frac{3x}{2} \right)^{2/3} - \Bar{g_\Lambda} \left( \frac{3x}{2} \right)^{2} - \Bar{g_r}\left( \frac{3x}{2} \right)^{-2/3} - \Bar{g_s}\left( \frac{3x}{2} \right)^{-2} -E \right)} dx \right),
\label{theta}
\end{equation}
where $x_l$ and $x_r$ are the values of $x$ where the energy $E$ intercepts the potential $V(x)$ on the left and right sides, respectively.

Now, we compute tunneling probabilities $TP_{WKB}$ Eq. (\ref{tpwkb}), for the potential $V(x)$ Eq. (\ref{potential}). Here, we consider models with $\Bar{g_c} > 0,\Bar{g_\Lambda} > 0, \Bar{g_r} \geq 0$ and $\Bar{g_s} < 0$. Aiming at studying how these probabilities depend on those quantities and the energy $E$, we fix the coupling constants $\Bar{g_i}$'s with $i={c,\Lambda,r,s}$ and the energy $E$, except the one under investigation. This strategy is applied to each one of the coupling constants and the energy. When studying the $\Bar{g_i}$'s a fixed energy value was selected. The scale factor values where the energy $E$ intercepts the potential, $x_l$ and $x_r$, are also relevant. They must be determined in order to calculate the probabilities. 

\subsection{Studying the tunneling probabilities from the scale factor origin}

The decision of choosing $g_s < 0$ implies on the presence of a positive infinite tall barrier at the origin of the potential. We decided that an investigation should be done about the tunneling probability associated to that tall barrier. To investigate the $TP_{WKB}$ behavior at the origin, we start by computing $\theta$ from equation (\ref{theta}). The chosen value of $E$ will determine the values of $x_l$ and $x_r$. We performed this calculation for different values of $E$, using a potential which has the shape shown in Figure \ref{potential-1}. All the calculations returned the same result saying that $\theta$ always diverges and goes to infinity. This implies on a null $TP_{WKB}$.  
This means that the Universe described in this model does not start from a zero scale factor.  
Thus, we shall assume, for the present model, that the Universe was initially formed in the region between the infinite tall barrier and the second barrier, where is located the well. Then, it has, initially, a small but non-zero size and is free from the big bang singularity, also, at the quantum level. We, also, assume that the universe was formed with an energy $E$ that is smaller than the second barrier maximum. A classical universe would stay in that region forever. Assuming that the universe was quantum mechanical, it was able to tunnel through the second barrier and begin its expansion. Here, we consider that the Universe is born right after it emerges from the second barrier. Taking in account these assumptions, we want to investigate, now, what is the probability that the universe tunnel through the second barrier and starts to expand. In the following Subsections, we compute the $TP_{WKB}$ for that tunneling process.

\subsection{$\Bar{g_c}$}

\begin{figure}[!h]
  \centering
	\includegraphics[width=0.45\textwidth]{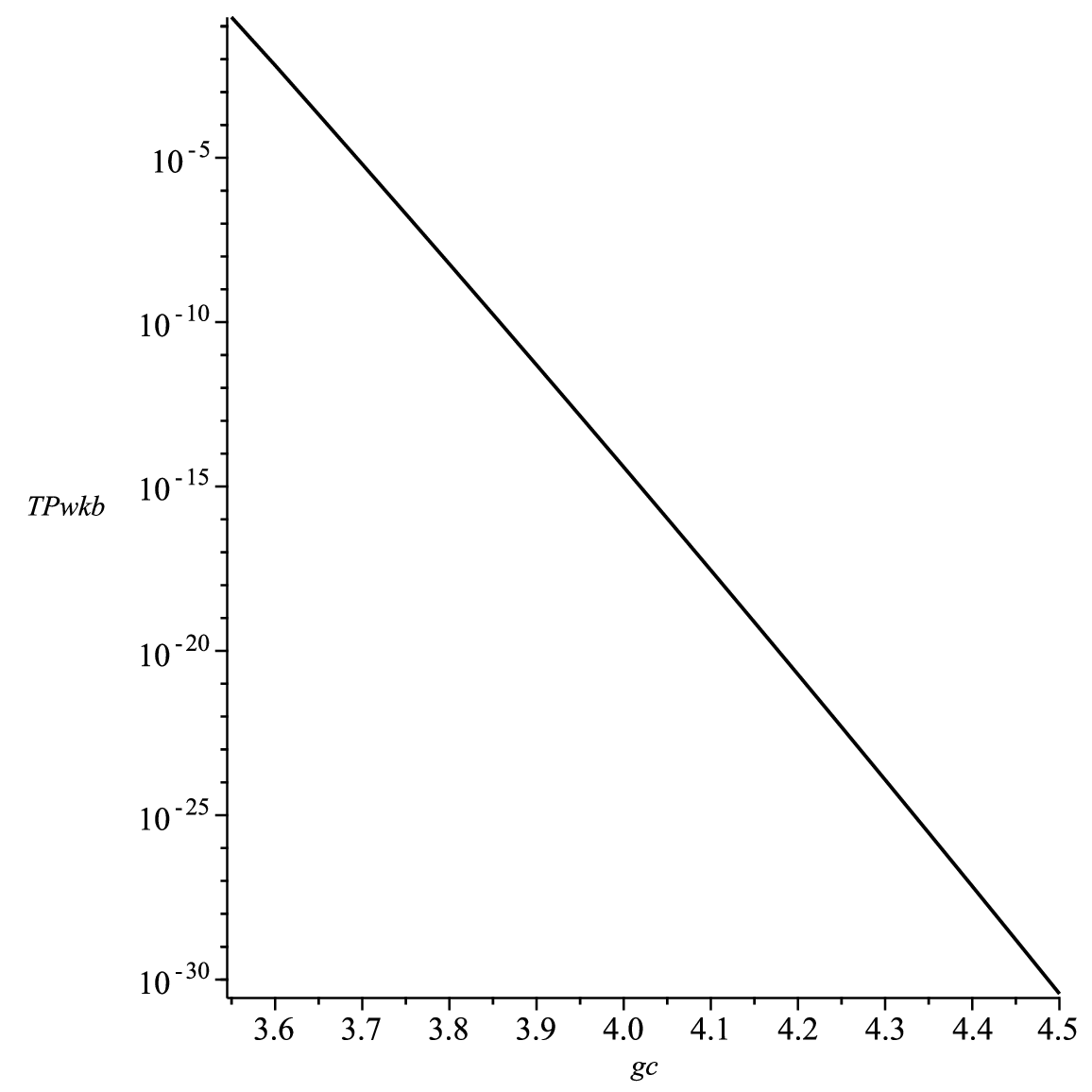}
  \caption{Variation of $TP_{WKB}$ as $\Bar{g_c}$ changes for a fixed energy $E = 8$.}
\label{var_gc}
\end{figure}

We start studying how $TP_{WKB}$ depends on the Hořava–Lifshitz coupling constant $\Bar{g_c}$. Then, we fix $\Bar{g_\Lambda}, \Bar{g_r}, \Bar{g_s}$ and $E$ and choose many different values of $\Bar{g_c}$. We calculate the tunneling probabilities for those many different values and conclude that $TP_{WKB}$ decreases as one raises $\Bar{g_c}$. So, it is expected that the universe is born with the smallest possible value of that coupling constant. Here, it is shown an example where we have a fixed energy $E = 8$ and the coupling constants are $\Bar{g_\Lambda} = 0.1, \Bar{g_r} = 0.2$ and $\Bar{g_s} = -0.001$. We compute the probabilities for 20 values of $\Bar{g_c}$, starting at $\Bar{g_c} = 3.55$ and finishing at $\Bar{g_c} = 4.5$, in steps of $0.05$. The maximum value of $V(x)$ is greater than the selected energy value for all values assumed by $\Bar{g_c}$. Figure \ref{var_gc} shows the curve $TP_{WKB} \times \Bar{g_c}$. 

\subsection{$\Bar{g_\Lambda}$}

\begin{figure}[!h]
  \centering
	\includegraphics[width=0.45\textwidth]{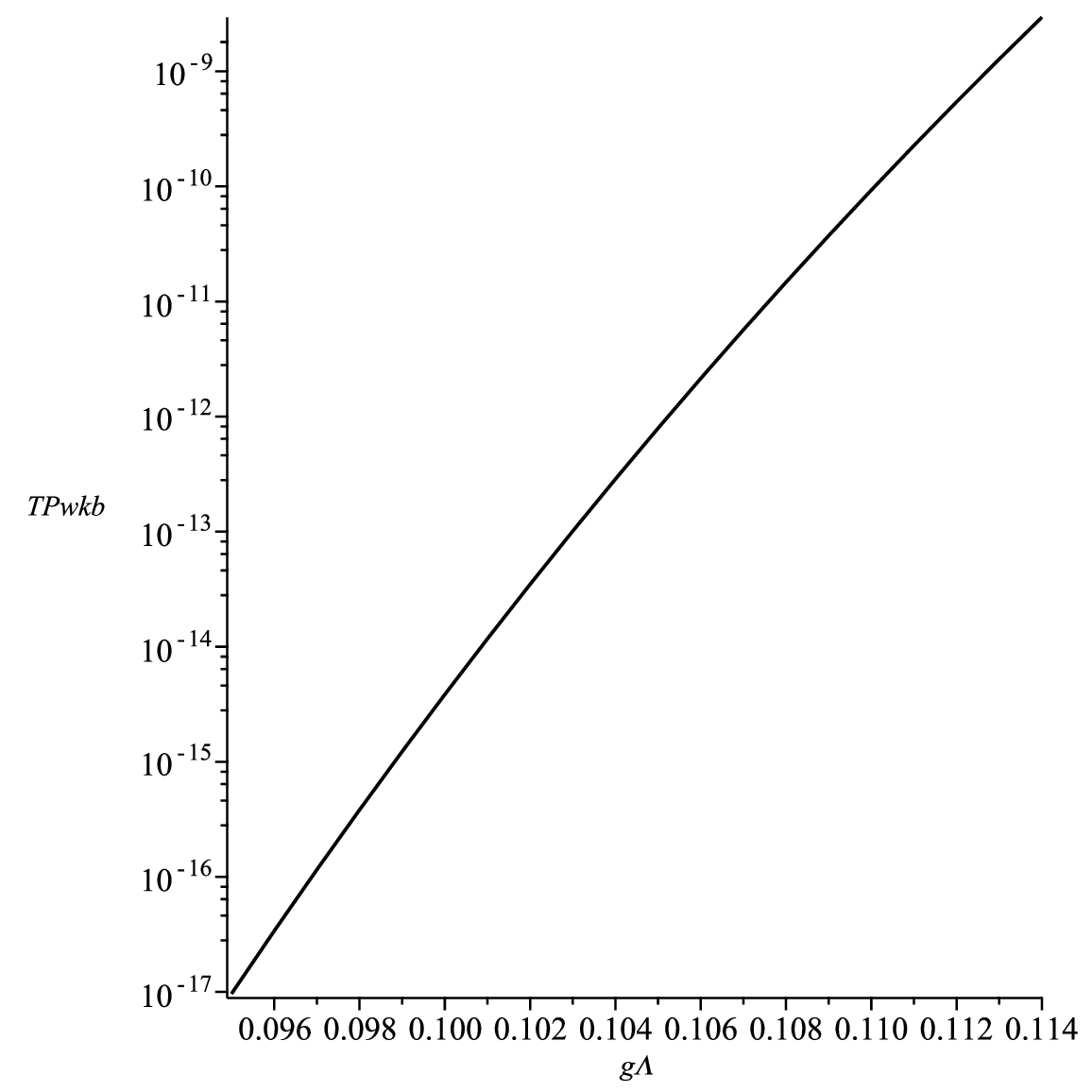}
  \caption{Variation of $TP_{WKB}$ as $\Bar{g_\Lambda}$ changes for a fixed energy $E = 8$.}
\label{var_glambda}
\end{figure}

In the present subsection, we study how $TP_{WKB}$ depends on $\Bar{g_\Lambda}$. We fix $\Bar{g_c}, \Bar{g_r}, \Bar{g_s}$ and $E$ and choose many different values of $\Bar{g_\Lambda}$. We calculate the tunneling probabilities for those many different values and conclude that $TP_{WKB}$ grows as one raises $\Bar{g_\Lambda}$. So, it is expected that the universe is born with the biggest possible value of that coupling constant. Here, it is shown an example where we have a fixed energy $E = 8$ and the coupling constants are $\Bar{g_c} = 4.0, \Bar{g_r} = 0.2$ and $\Bar{g_s} = -0.001$. We compute the probabilities for 20 values of $\Bar{g_\Lambda}$, starting at $\Bar{g_\Lambda} = 0.095$ and finishing at $\Bar{g_\Lambda} = 0.114$, in steps of $0.001$. The maximum value of $V(x)$ is greater than the selected energy value for all values assumed by $\Bar{g_\Lambda}$. Figure \ref{var_glambda} shows the curve $TP_{WKB} \times \Bar{g_\Lambda}$.

\subsection{$\Bar{g_r}$}

\begin{figure}[!h]
  \centering
	\includegraphics[width=0.45\textwidth]{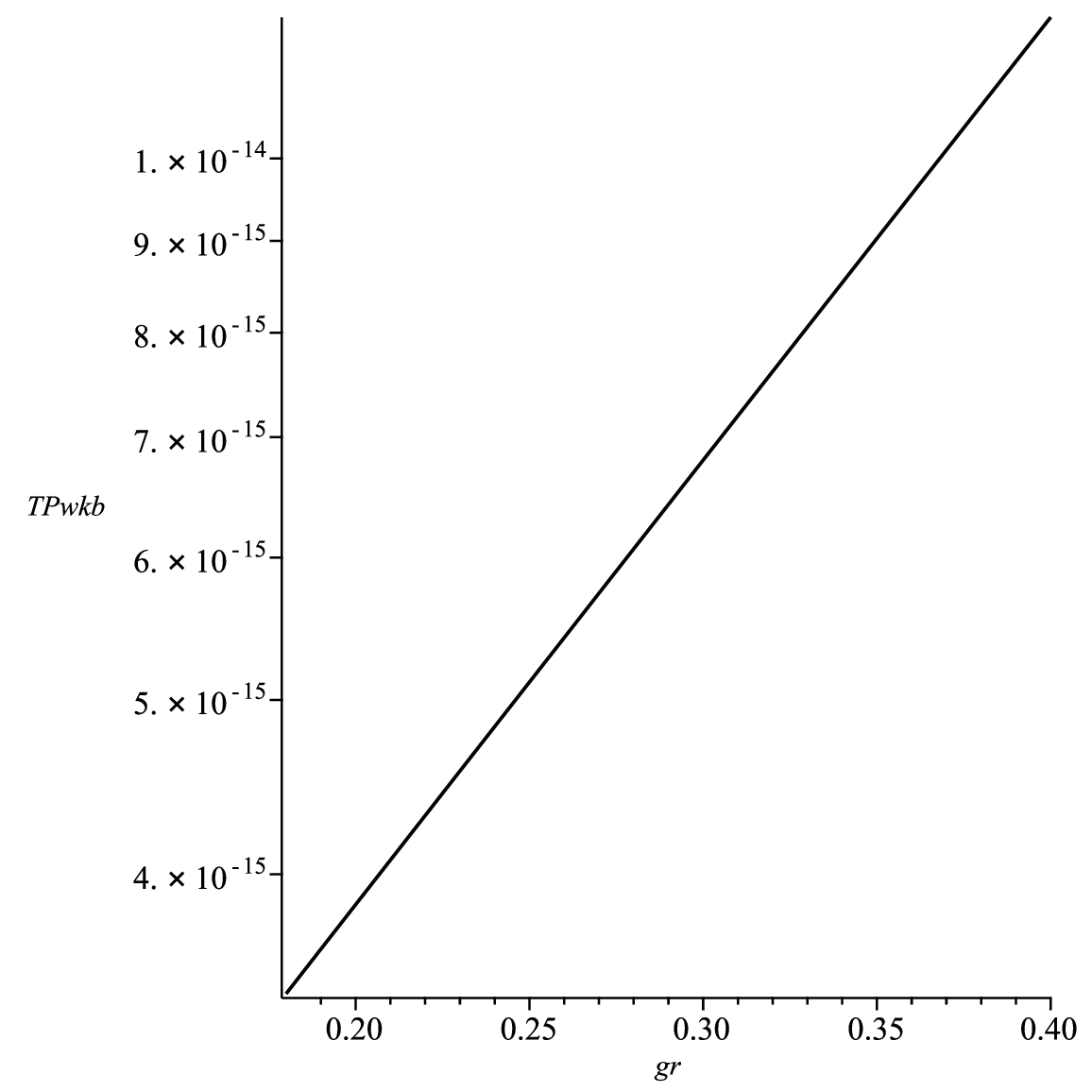}
  \caption{Variation of $TP_{WKB}$ as $\Bar{g_r}$ changes for a fixed energy $E = 8$.}
\label{var_gr}
\end{figure}

In the present subsection, we study how $TP_{WKB}$ depends on $\Bar{g_r}$. We fix $\Bar{g_c}, \Bar{g_\Lambda}, \Bar{g_s}$ and $E$ and choose many different values of $\Bar{g_r}$. We calculate the tunneling probabilities for those many different values and conclude that $TP_{WKB}$ increases as one raises $\Bar{g_r}$. So, it is expected that the universe is born with the biggest possible value of that coupling constant. Here, it is shown an example where we have a fixed energy $E = 8$ and the coupling constants are $\Bar{g_c} = 4.0, \Bar{g_\Lambda} = 0.1$ and $\Bar{g_s} = -0.0001$. We compute the probabilities for 23 values of $\Bar{g_r}$, starting at $\Bar{g_r} = 0.18$ and finishing at $\Bar{g_r} = 0.40$, in steps of $0.01$. The maximum value of $V(x)$ is greater than the selected energy value for all the values assumed by $\Bar{g_r}$. Figure \ref{var_gr} shows the curve $TP_{WKB} \times \Bar{g_r}$. 

\subsection{$\Bar{g_s}$}

\begin{figure}[!h]
  \centering
	\includegraphics[width=0.45\textwidth]{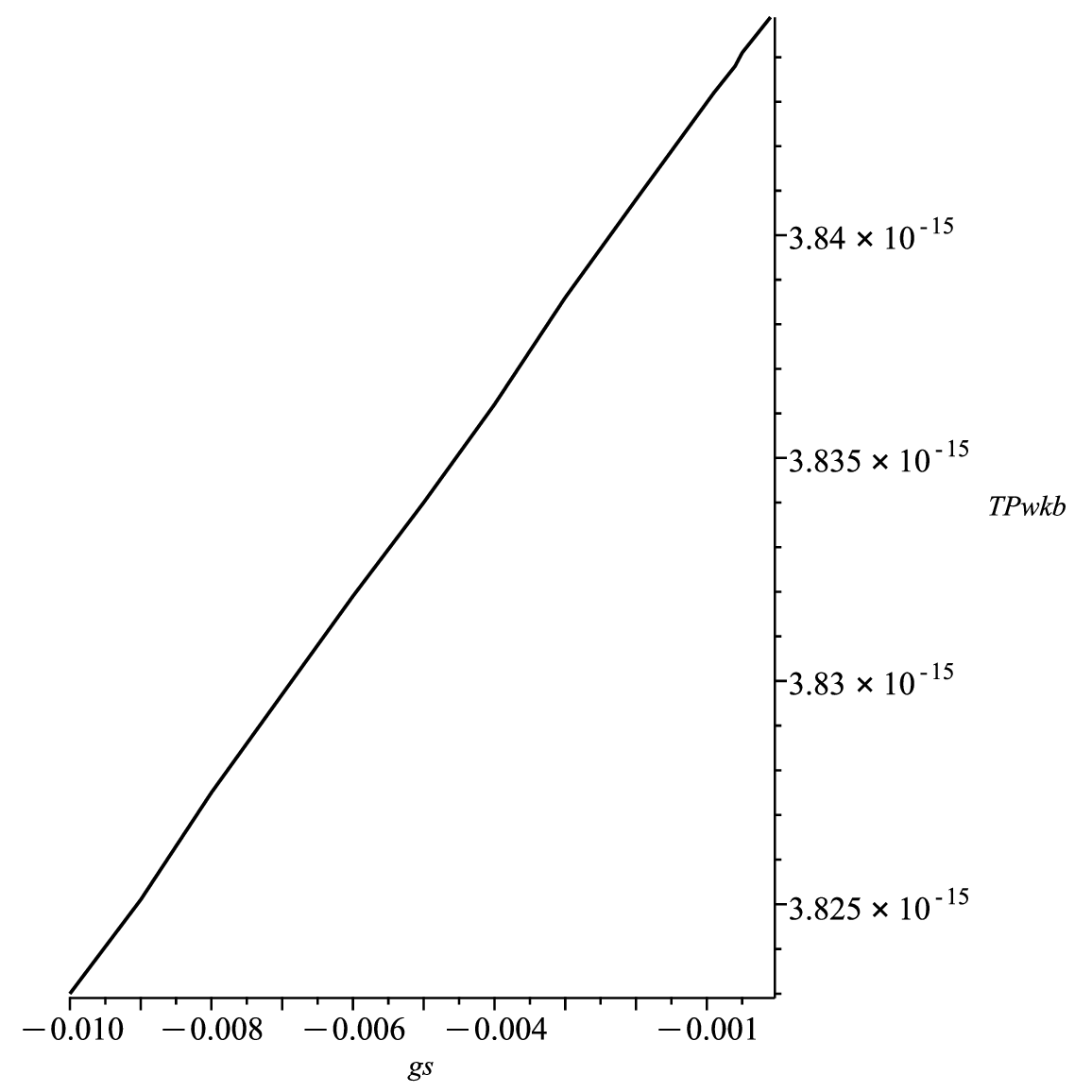}
  \caption{Variation of $TP_{WKB}$ as $\Bar{g_s}$ changes for a fixed energy $E = 8$.}
\label{var_gs}
\end{figure}

In the present subsection, we study how $TP_{WKB}$ depends on $\Bar{g_s}$. We fix $\Bar{g_c}, \Bar{g_\Lambda}, \Bar{g_r}$ and $E$ and choose many different values of $\Bar{g_s}$, using only negative values. We calculate the tunneling probabilities for those many different values and conclude that $TP_{WKB}$ increases as one raises $\Bar{g_s}$. So, it is expected that the universe is born with the biggest possible value of that coupling constant. Here, it is shown an example where we have a fixed energy $E = 8$ and the coupling constants are $\Bar{g_c} = 4.0, \Bar{g_\Lambda} = 0.1$ and $\Bar{g_r} = 0.2$. We compute the probabilities for 18 values of $\Bar{g_s}$, dividing the $\Bar{g_s}$ interval in two parts. The first part starts at $\Bar{g_s} = -0.01$ and finishes at $\Bar{g_s} = -0.001$, in steps of $0.001$. The second part starts at $\Bar{g_s} = -0.001$ and finishes at $\Bar{g_s} = -0.0001$, in steps of $0.0001$. The maximum value of $V(x)$ is greater than the selected energy value for all the values assumed by $\Bar{g_s}$. Figure \ref{var_gs} shows the curve $TP_{WKB} \times \Bar{g_s}$. 

\subsection{Energy}

\begin{figure}[!h]
  \centering
	\includegraphics[width=0.45\textwidth]{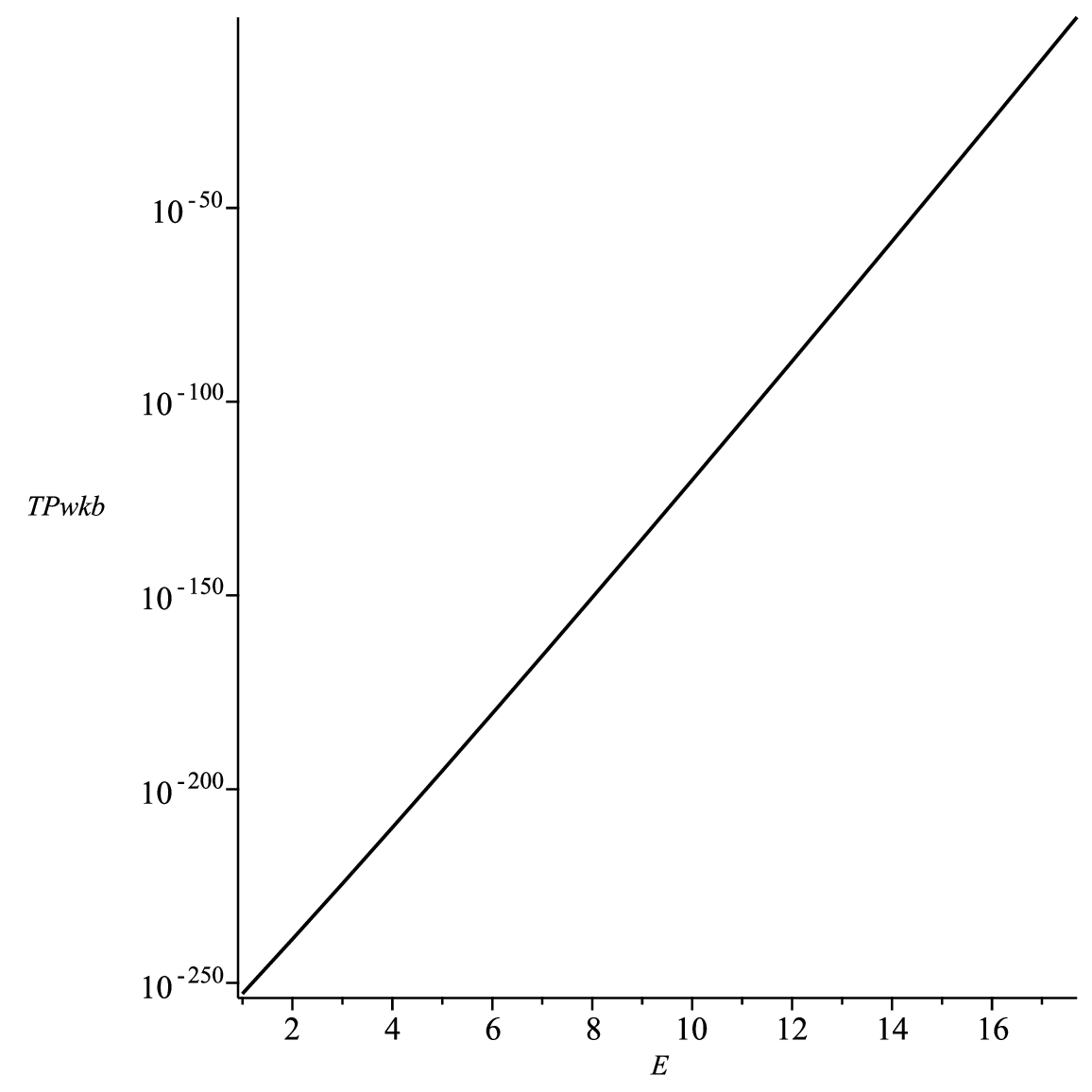}
  \caption{Variation of $TP_{WKB}$ as one varies the energy $E$.}
\label{var_E}
\end{figure}

In the present subsection, we study how $TP_{WKB}$ depends on the fluid energy $E$. We fix the HL's coupling constants $\Bar{g_c}, \Bar{g_\Lambda}, \Bar{g_r}$ and $\Bar{g_s}$ and choose many different values of $E$, using only positive values. We calculate the tunneling probabilities for those many different values and conclude that $TP_{WKB}$ increase as one raises $E$. So it is expected  that the Universe will be born with the biggest possible value of the fluid energy $E$. Here, it is shown an example where we have fixed the coupling constants as $\Bar{g_c} = 4.0, \Bar{g_\Lambda} = 0.03$, $\Bar{g_r} = 0.2$ and $\Bar{g_s} = -0.001$. We compute the probabilities for 19 values of $E$, starting at $E = 1$ and finishing at $E = 17$, in steps of $1$. In addition, we compute $TP_{WKB}$ for two other values, $E = 17.5$ and $E = 17.7$. The maximum value of $V(x)$ is greater than all the values of $E$ used in this study. Figure \ref{var_E} shows the curve $TP_{WKB} \times E$.

\section{Conclusion}

In this letter, we applied quantum cosmology to investigate the early moments of a FLRW cosmological model, 
using HL as the gravitational theory. The matter content of the model is a dust perfect fluid and the spatial sections have positive curvature.
Regarding the HL theory, we considered the {\it projectable} version of that theory without the {\it detailed balance condition}.
We started studying the classical model and obtained its total Hamiltonian. Imposing some conditions on the HL's coupling constants, we obtained the most general potential barrier from $V(x)$ Eq. (\ref{potential}). It has the shape of an infinite tall barrier at the origin, followed by a well, followed by a second barrier. The choice of $\Bar{g_s} < 0$ implies that the potential has the infinite tall barrier at the origin and those models are free from the {\it big bang} singularity, at the classical level. From that total Hamiltonian, we draw the phase portrait of the model and from it we identified, qualitatively, all possible scale factor classical trajectories. Then, we 
quantized the model and found the appropriate Wheeler-DeWitt equation. In order to avoid factor ordering ambiguities, in the Wheeler-DeWitt equation, we introduced a canonical transformation. We solved that equation using the WKB approximation and computed the tunneling probabilities for the birth of that universe ($TP_{WKB}$). With the aid of the $TP_{WKB}$, we concluded that, in the present model, the universe was initially formed in the region between the infinite tall barrier and the second barrier, due to the presence of the tall barrier at the origin. Then, it has, initially, a small but non-zero size and is free from the {\it big bang} singularity, also, at the quantum level.
Since, we considered that the Universe is born right after it emerges from the second barrier, we computed how $TP_{WKB}$ depends on $E$ and all HL's coupling constants. The result of these computations were the following: (i) $TP_{WKB}$ decrease when one increases $\Bar{g_c}$; and (ii) $TP_{WKB}$ grows when one increases $\Bar{g_\Lambda}$, $\Bar{g_r}$, $\Bar{g_s}$ and $E$. Therefore, the universe should starts with the greatest possible values of $\Bar{g_\Lambda}$, $\Bar{g_r}$, $\Bar{g_s}$ and $E$ and the smallest possible value of $\Bar{g_c}$.

{\bf Acknowledgments}. A. Oliveira Castro Júnior thanks Coordena\c{c}\~{a}o de\\ Aperfei\c{c}oamento de Pessoal de N\'{i}vel Superior (CAPES).
G. A. Monerat thanks FAPERJ for financial support and Universidade do Estado do Rio de Janeiro (UERJ) for the Proci\^{e}ncia grant.

{\bf Data availability statement}. All data generated or analyzed during this study are included in this published article.


\end{document}